\documentclass[a4paper,11pt]{article}
\usepackage{jinstpub} % for details on the use of the package, please see the JINST-author-manual
\usepackage{lineno}

\title{\boldmath Demonstrator System Testing and Performance for the ATLAS ITk Pixel Detector for HL-LHC}

\usepackage{natbib}

\author{Y. Khwaira on behalf of the ATLAS ITK Pixel Group}
\affiliation{
Laboratoire de physique nucléaire et des hautes énergies (LPNHE) - CNRS - Sorbonne Université/Université Paris Cité,\\
4 Place Jussieu, Tower 22, 1st floor, 75005 Paris
}

\emailAdd{ykhwaira@cern.ch}

\abstract{A demonstrator for each slice of the ATLAS pixel detector was built to replicate the real detector and provide early solutions for operating and maintaining its components. This system-level testing of the all-silicon Inner Tracker (ITk) pixel detector for the ATLAS experiment at CERN's HL-LHC encompasses a wide array of system components, which is essential for managing the increased luminosity and radiation levels expected at HL-LHC, thereby enhancing tracking performance. Utilizing advanced silicon sensor technologies, serial powering, and lightweight carbon fiber structures, the demonstrator and assembled components on the support structure will undergo several studies for verification and commissioning. Extensive tests on serial powering, monitoring, and data acquisition were conducted, ensuring the system’s robustness and reliability for future high-energy physics experiments. Additionally, three different sub-components will be introduced for the novel ITk pixel detector, specifically designed for the outer barrel (OB), outer end caps (OEC), and inner system (IS) sections.
}


\arxivnumber{2411.06992} % Only if you have one

\begin{document}
\maketitle
\flushbottom

\section{Introduction}
\label{sec:intro}

The integrated luminosity of the LHC will be increased tenfold over its original design threshold, for the transition into the High Luminosity-LHC (HL-LHC) \cite{ZurbanoFernandez:2020cco}. This will significantly increase the LHC collision reach, leading to higher detector occupancy and accelerated radiation damage. To address these challenges, the current detectors must be replaced and upgraded to cope with the new requirements and demands of the HL-LHC, with faster readout detectors, and with higher granularity. Specifically, the ATLAS experiment \cite{TheATLASCollaboration2008} will upgrade its Inner Detector (ID) to a new all-silicon Inner Tracker (ITk) system  \cite{Meng:2021jes}.

The ITk comprises of a four-layer strip detector with planar sensors and a five-layer hybrid pixel detector which is the focus of our paper, covering approximately 13 m² and featuring around 10,000 hybrid modules. The inner layer is equipped with 3D sensors, while planar sensors are deployed on the outer layers. This new design promises enhanced performance but also presents complex operational demands, requiring extensive system testing and integration studies to mitigate the system complexities.

Three demonstrator systems are under development as proof-of-concept platforms to validate the full ITk pixel detector production workflow. These demonstrators integrate key subsystems, including silicon sensors with RD53A quad modules, on- and off-detector services, multi-module readout, signal aggregation, optical conversion over a 65 m fiber and trunk system, and operation of detector control systems. This paper presents the pixel system, which interfaces the complete detector system, covering on- and off-detector services, data transmission, DAQ/DCS, integration, common electronics, and module-based RD53A quads.

\section{ Demonstrator Setup and Associated Sub-systems}

\begin{figure}[htbp]
    \centering
    \includegraphics[width=0.7\linewidth]{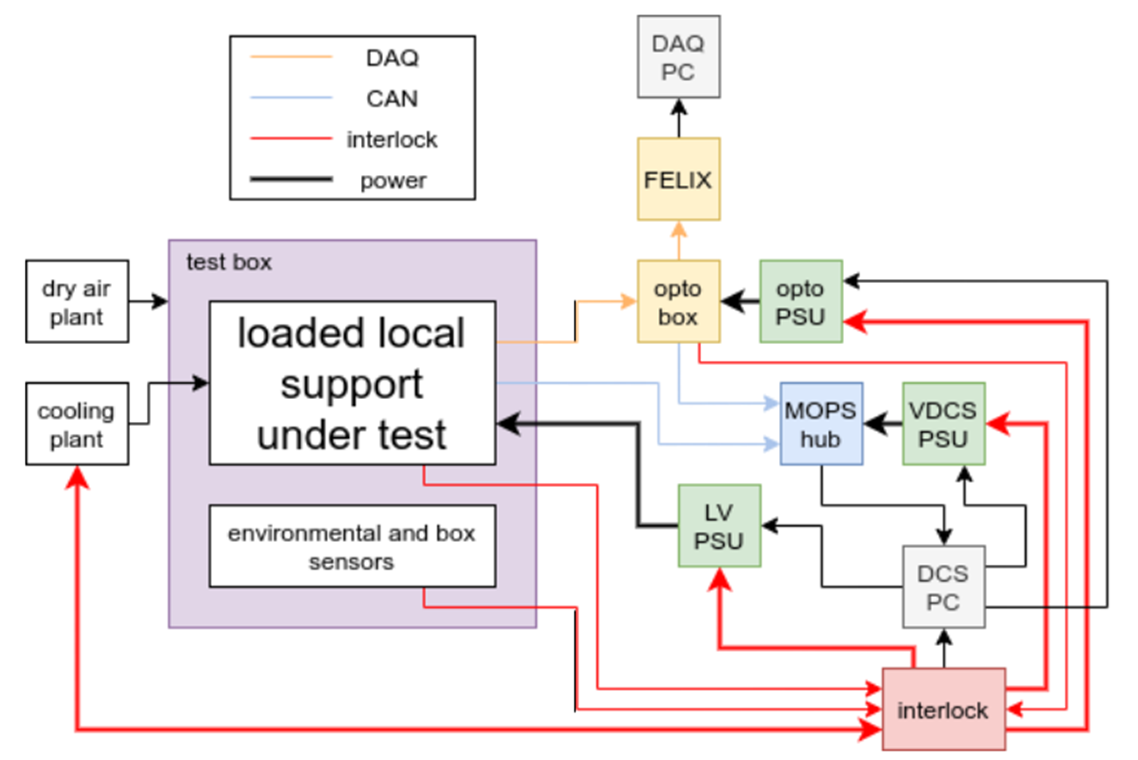}
    \caption{Overview of the main on/off detector systems for the OB demonstrator testing setup.}
    \label{fig:overview_of_demonstrator_setup__}
\end{figure}

The demonstrator setup is built around two interconnected systems as Figure \ref{fig:overview_of_demonstrator_setup__} illustrates: on-detector and off-detector services. The on-detector services include module power supplies, a  \texorpdfstring{$CO_2$}{CO2} evaporative cooling plant for efficient cooling and a dedicated Monitoring of Pixel System (MOPS) \cite{Walsemann:20203c} for monitoring operational status, and for tracking individual power channels and module temperatures. Serial powering groups modules into chains powered by a constant current, with Shunt Low Dropout (SLDO) regulators for shunting excess current.

The off-detector services focus on realistic cabling—Type-1 bundles, patch panels, Type-3 LV via serial powering to reduce cabling, HV, Detector Control System (DCS) \cite{Joos:2022}, and interlock cables. Data transmission and command signals are managed by electrical services interfacing modules with the Opto-box for optical conversion and signal aggregation outside the test box for radiation safety and accessibility as in Figure \ref{fig:overview_of_demonstrator_setup__}.

Testing requires realistic cable length with a setup optimized in a lightproof, temperature-controlled test box that supports electrical and thermal tests, prevents humidity, and includes sensors for monitoring key parameters. Furthermore, DCS manages environmental sensors, cooling, interlock crate, MOPS, Opto-box \cite{Guettouche:2022}, and loaded staves. The DAQ system, Front-End Link eXchange (FELIX) \cite{SolansSanchez:2020}, centralizes operations for the ITk pixel sensors.

The demonstrator's testing program includes thermal and electrical tests, with system checks for MOPS, serial powering, and cooling efficiency. These tests identify issues in integration, module performance, and NTC  temperature sensor consistency, ensuring the demonstrator’s readiness for large-scale production. 

\section{Readout Chain, DCS, and Data Transmission}
\label{sec:Section2_readout}

The demonstrator system serves as a proof of concept for the ITk design, showcasing and testing the solutions envisioned for the final installation. By minimizing cabling needs through serial powering and using electrical services to route data and command signals to Opto-converters, the demonstrator replicates key aspects of the intended system, providing critical insights into its implementation. Signal aggregation and electrical-to-optical conversion occur in the Opto-box outside the test environment for radiation shielding and component accessibility. Each quad module uplink is supported by 1.28 Gb/s fibers to manage high data volumes, and cable lengths reflect the real setup for bandwidth and transmission reliability testing.

Moreover, module readout is handled via the FELIX system to enable indirect, high-speed data transmission between on-detector and off-detector elements. It operates by sending configuration and control commands to front-end chips via fiber downlinks and aggregates data via fiber uplinks at speeds up to 10.24 Gb/s. E-link aggregation is managed by the lpGBT chip  \cite{Guettouche:2022}, ensuring efficient communication across modules. Optical signal transformations are facilitated by the VTRx+ \cite{Troska2018}
transceiver, assembled on Opto boards housed within the Opto-box.

Furthermore, the DCS  controls component powers and integrates it into a finite state machine (FSM) interface, simplifying setup operation and safety by the interlock. The FSM allows for quick navigation between component states via a graphical user interface (GUI), enabling efficient operation for the ITk pixel detector's future deployment.

\begin{figure}[htbp]
    \centering
    \includegraphics[width=0.75\linewidth]{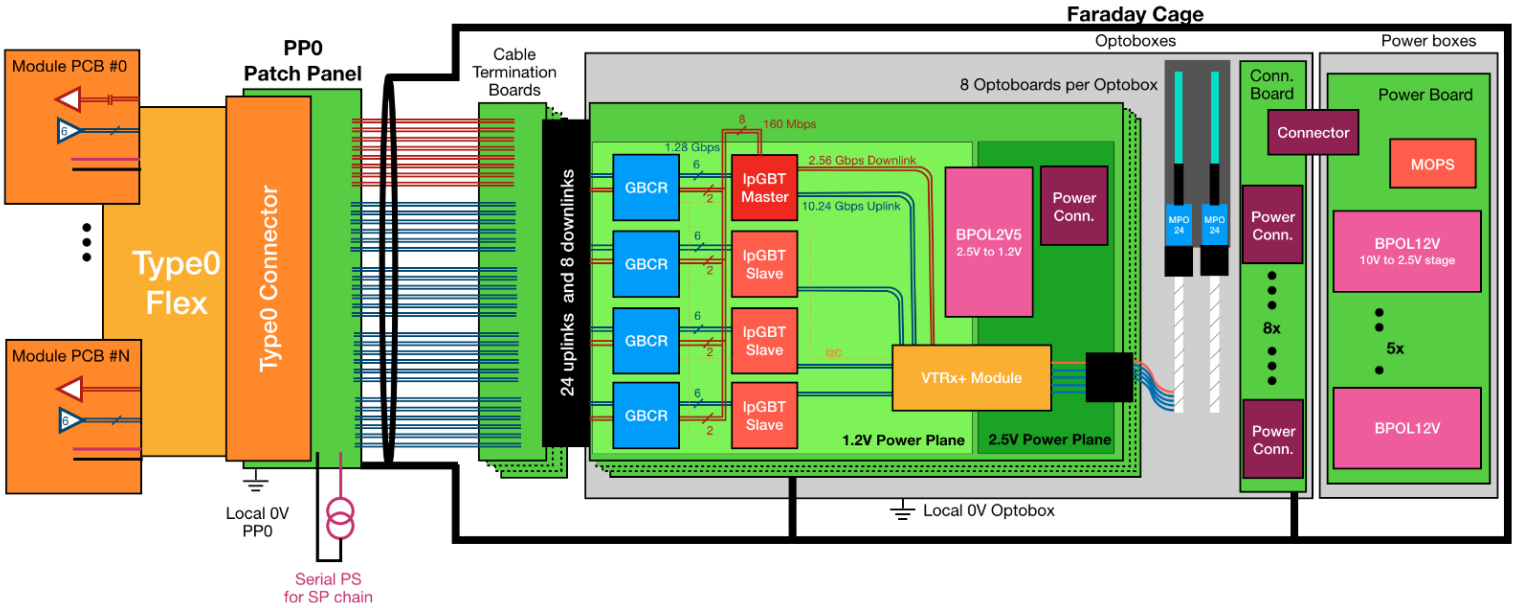}
    \caption{Complete readout setup with the opto-system and data transmission for multiple modules.}
    \label{fig:Opto_box_label}
\end{figure}

\section{Local Loaded Supports and  \texorpdfstring{$CO_2$}{CO2} Cooling System}
Mechanical structures play a vital role in stabilizing module positions relative to the global coordinates of the overall detector volume, providing essential structural support near the interaction point (IP) around the beam pipe. 

%%Moreover, it is an important key to maintaining optimal thermal and mechanical properties. These supports rely on carbon fiber-reinforced plastic (CFRP) for its minimal thermo-elastic stress response, structural stiffness, and ease of handling under variable thermal conditions.

\begin{figure}[htbp]
\centering
\includegraphics[width=.42\textwidth]{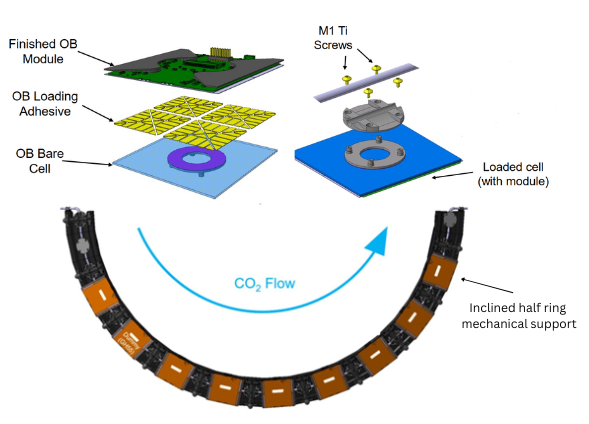}
\qquad
\includegraphics[width=.52\textwidth]{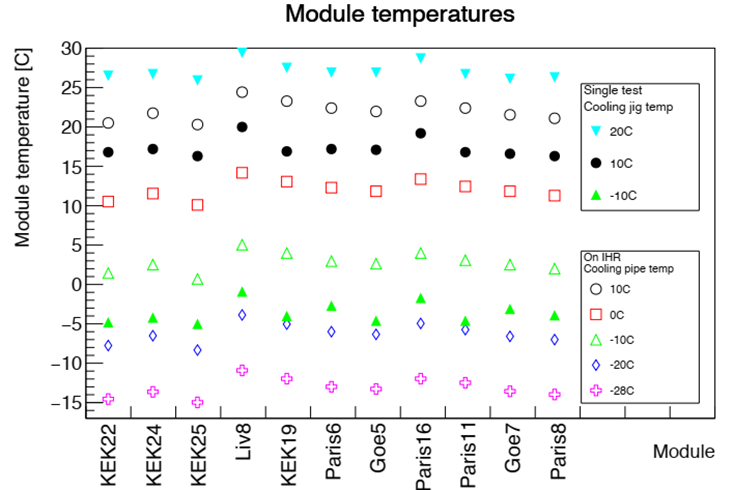}
\caption{ Figure (left) shows  \texorpdfstring{$CO_2$}{CO2} flow from inlet to exhaust with the module added to the cell-loading structure on the stave. Figure (right) shows thermal variations in NTC measurements that were observed between single-module cooling jig tests and via the \texorpdfstring{$CO_2$}{CO2} cooling pipe.}
\label{fig:CO2_cooling_and_mechanical_supp}
\end{figure}

Precision in assembly is critical, especially regarding the layer of glue used to attach modules to the stave \cite{MunozSanchez2023}. Excessive glue can shift module positions and reduce the thermal interface, negatively impacting heat dissipation from modules to the cooling system. Each stave integrates three key elements for optimized thermal management: thin-walled titanium tubes that circulate  \texorpdfstring{$CO_2$}{CO2} coolant, base cooling blocks strategically placed to create thermally conductive interfaces, and carbon-fiber composites that form the primary support structure as Figure \ref{fig:CO2_cooling_and_mechanical_supp} explicitly shows, ensuring stability while managing temperature fluctuations.

Testing the cooling efficiency across modules both before and after integration is crucial to maintain consistent thermal performance. Comparative studies shown in Figure \ref{fig:CO2_cooling_and_mechanical_supp} were conducted on module temperatures in a dedicated cooling jig and on the demonstrator setup, assessing variations at multiple  \texorpdfstring{$CO_2$}{CO2} temperature set-points. Initial results from MOPS indicated slight offsets, averaging around 5 °C, consistent across different modules in the demonstrator setup. The  \texorpdfstring{$CO_2$}{CO2} system, using serial cooling across modules, maintained stable performance, effectively reproducing the expected cooling behavior and confirming the robust design of both the loaded supports and cooling system for reliable operation under HL-LHC conditions.

\section{RD53A Module Reliability in Full-System Tests}

\begin{figure}[htbp]
    \centering
    \includegraphics[width=0.75\linewidth]{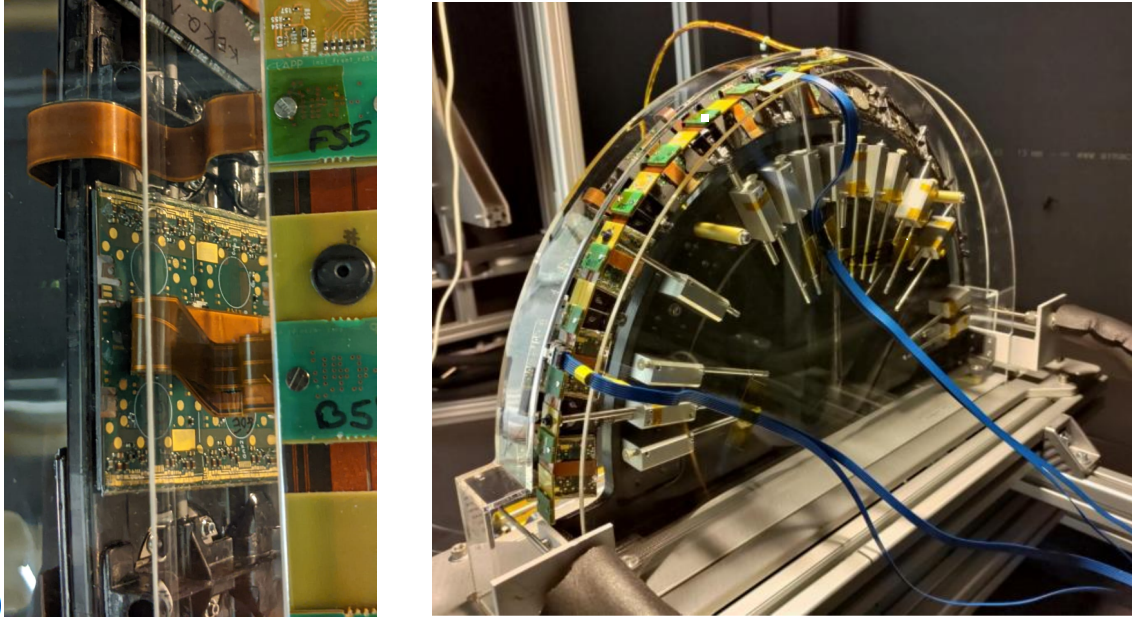}
    \caption{Inclined Half-Ring (IHR) stave installed inside the test box and mounted with 10 cell-loaded RD53A quad modules attached with pig-tail cables via the patch panel 0 (PP0).}
    \label{fig:IHR_loaded_with_modules}
\end{figure}

\begin{figure}[htbp]
    \centering
    \includegraphics[width=1\linewidth]{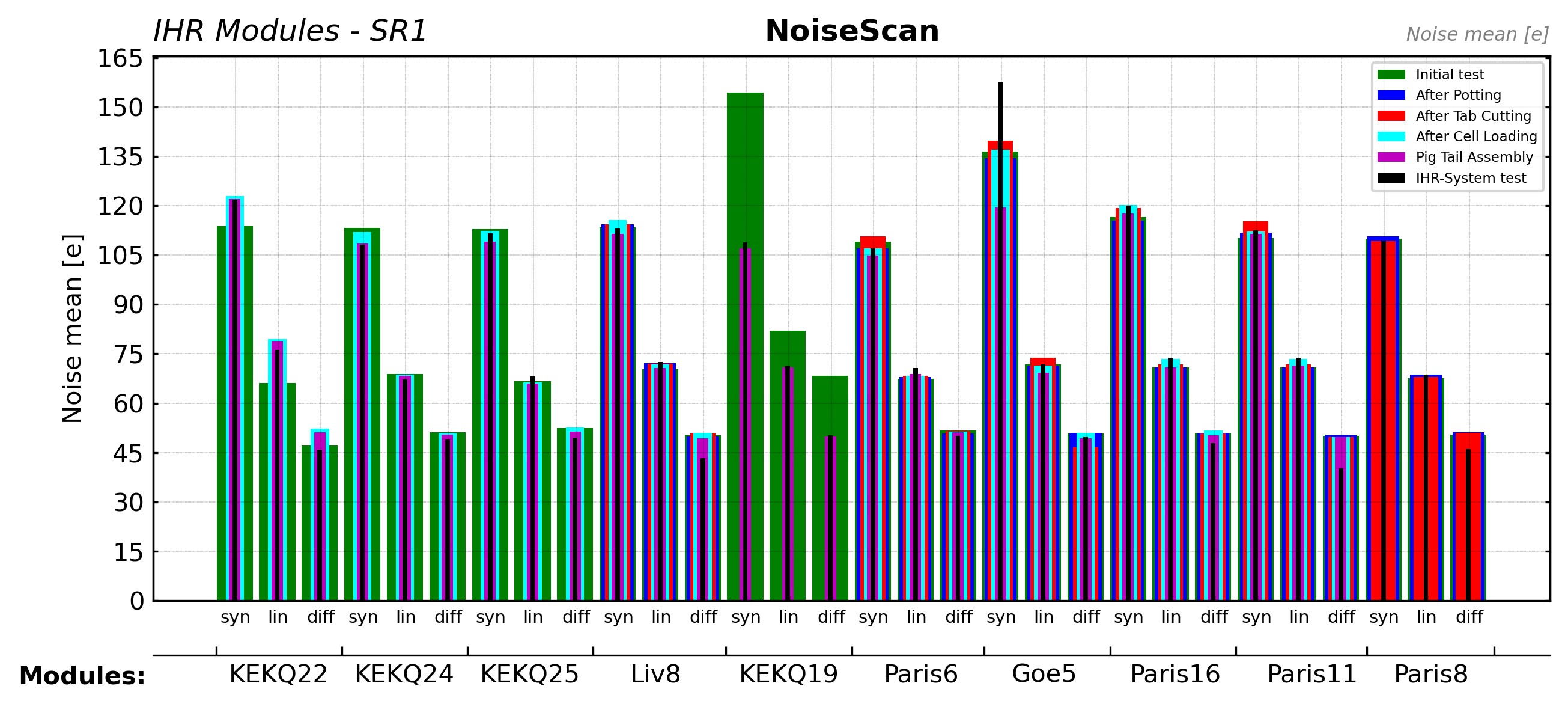}
    \caption{The figure illustrates the multiple testing stages that RD53a modules undergo throughout various production steps. Upon arrival at CERN, an initial test is conducted to evaluate baseline functionality. Then, wire bonding is secured by coating for protection. The modules then proceed to cell loading on the back of the module as shown in Figure \ref{fig:CO2_cooling_and_mechanical_supp} and pig-tail assembly before entering the final OB demonstrator system test phase \cite{Khwaira:2883908}. Notably, noise scan performance remains consistent across all stages, indicating stable behavior for the IHR modules in the OB demonstrator.}
    \label{fig:Noise_scan_for_the_IHR}
\end{figure}

A key objective of the OB demonstrator project is to validate the loading concepts crucial to ITk detector construction with the overall sub-systems described in section \ref{sec:Section2_readout}. Using detector prototypes equipped with RD53A electronics \cite{Loddo:2024teb} as in  Figure \ref{fig:IHR_loaded_with_modules}, the electrical scans taken at each stage of the production flow ensure the reliability of the module’s performance. A large 
set of electrical scans are needed to characterize the functionality of individual module components using standardized quality control (QC) measures across all production stages. To ensure direct inspections of any potential degradation by comparing the output to the previous and following
stages. A final round of electrical scans evaluates the fully loaded modules in the ITk OB demonstrator setup, as illustrates 10 modules loaded on the IHR in Figure \ref{fig:Noise_scan_for_the_IHR} of the OB demonstrator at SR1 in CERN. Which features the full-system test of DAQ,  \texorpdfstring{$CO_2$}{CO2} cooling, SLDO serial powering chains, and collective module testing. The performance across all six testing stages demonstrates consistency, with minimal variation across stages despite environmental differences and fundamentally different setups.

%%\begin{figure}[htbp]
    %%\centering
    %%\includegraphics[width=1\linewidth]%%{sample_of_ihr_results.PNG}
    %%\caption{Summary of all QC production stages for Goe5 %%chip 4 indicating a well-preserved performance.}
 %%   \label{fig:Goe4_chip_perfromance}
%%\end{figure}

\section{Conclusion}
The demonstrator setup is instrumental in validating key ITk system aspects and proofs that the modules across the production demonstrate similar and consistent performance. Nevertheless, the demonstrator assisted in validating the loaded local support structure, services, and overall detector design for the final design review (FDR). Moreover, serial powering and grounding tests on the RD53A quad modules with commissioning studies, including successful electrical scans with the complete readout path, in addition to the  \texorpdfstring{$CO_2$}{CO2} cooling plant facilitated the progression to system tests.

\hfill

\paragraph{Note:} Copyright 2024 CERN for the benefit of the ATLAS Collaboration. Reproduction of this article or parts of it is allowed as specified in the CC-BY-4.0 license.

% Bibliography

%% [A] Recommended: using JHEP.bst file
%% \bibliographystyle{JHEP}
%% \bibliography{biblio.bib}

%% or
%% [B] Manual formatting (see below)
%% (i) We suggest to always provide author, title and journal data or doi:
%% in short all the informations that clearly identify a document.
%% (ii) please avoid comments such as "For a review'', "For some examples",
%% "and references therein" or move them in the text. In general, please leave only references in the bibliography and move all
%% accessory text in footnotes.
%% (iii) Also, please have only one work for each \bibitem.

\end{document}